\documentstyle[epsf,epsfig,twocolumn,prl,aps]{revtex}

\newcommand{\be}{\begin{equation}}
\newcommand{\ee}{\end{equation}}
\newcommand{\bea}{\begin{eqnarray}}
\newcommand{\eea}{\end{eqnarray}}

\newcommand{\gton}{\stackrel{>}{\sim}}
\newcommand{\lton}{\mathrel{\lower.9ex
                  \hbox{$\stackrel{\displaystyle <}{\sim}$}}}

\begin{document}

\title{The $K/\pi$ ratio from condensed Polyakov loops}
\author{O.\ Scavenius$^a$, A.\ Dumitru$^b$, J.T.\ Lenaghan$^c$}
\address{
$^a$ NORDITA, Blegdamsvej 17, DK-2100 Copenhagen {\O}, Denmark\\
$^b$ Department of Physics, Brookhaven National Laboratory, 
Upton, NY 11973, USA \\
$^c$ The Niels Bohr Institute, Blegdamsvej 17, 
DK-2100 Copenhagen {\O}, Denmark}

\maketitle   

\begin{abstract}
We perform a field-theoretical computation of hadron production 
in large systems at the QCD
confinement phase transition associated with restoration of
the $Z(3)$ global symmetry. This occurs from the
decay of a condensate for the Polyakov loop. 
From the effective potential for the Polyakov loop, its mass just below
the confinement temperature $T_c$ is in between the vacuum masses of the
pion and that of the kaon. Therefore, due to phase-space restrictions the
number of produced kaons is roughly an order of magnitude smaller than that
of produced pions, in agreement with recent results from collisions of gold
ions at the BNL-RHIC. From its mass,
we estimate that the Polyakov loop condensate 
is characterized by a (spatial) correlation scale of
$1/m_\ell\simeq 1/2$~fm. For systems of 
deconfined matter of about that size, the free energy 
may not be dominated by a condensate for the Polyakov loop, and 
so the process of hadronization may be qualitatively different as
compared to large systems. In that vein, experimental data on hadron abundance
ratios, for example $K/\pi$, in high-multiplicity $pp$ events at high energies
should be very interesting.

\end{abstract}

\narrowtext      


High-energy inelastic processes provide the unique opportunity to 
recreate perhaps the high energy density state of QCD matter that presumably
prevailed during the first microseconds of the evolution of the early
universe: the so-called Quark-Gluon Plasma (QGP).
Simply speaking, the QGP is the state of matter for which quarks and gluons are
deconfined and chiral symmetry is restored.
Even if produced, the QGP is a transient state of matter which can not be
observed directly. Experimentally, one observes hadrons which are produced from
the decay of the deconfined QGP state. The hadron production process itself
is commonly called hadronization.

One might be able to gain some insight into the hadronization process by
studying the relative populations of various hadronic states emerging from
a high-energy inelastic process~\cite{Hage}.
In fact, one of the earliest proposed
signatures for a transient QGP state is the enhanced production
of strangeness, particularly the relative abundance of kaons to 
pions (the ``$K/\pi$ ratio'')~\cite{Rafelski:1982pu}. 
More recently, it has been discussed that a large variety of ratios of
hadronic multiplicities in high-energy heavy-ion collisions resemble those
of a hadron gas at a temperature $T_h$ close to
$T_c$, the confinement/deconfinement transition
temperature~\cite{thermo,thf,kabana,redlich}. This has been interpreted
such that the QGP state hadronizes statistically, in that all hadronic states
are populated according to their weight in the partition function of
a hadron gas~\cite{thf,stat,becattini}.
This interpretation was partly based also on the success of
the statistical hadronization model to describe relative hadron abundances
in high-energy inelastic $e^+e^-$ and $pp$, or $p\bar p$
reactions~\cite{becattini}. In such
reactions, one does not expect that a thermally and chemically
equilibrated QGP is formed prior to hadronization. Nevertheless, 
the hadronic final states appear to be
populated approximately according to a statistical distribution.
{}From $e^+e^-$ to central collisions of large nuclei,
the hadronization temperature obtained from fitting the relative
multiplicities was found to be more or less the same, and that it is also
independent of collision energy at high energies~\cite{becattini}.
This was interpreted~\cite{stat,becattini} as
a ``limiting temperature'' for hadron formation in inelastic reactions, which
is consistent with the prediction of a deconfinement temperature $T_c$
from lattice QCD above which hadronic states presumably can not freeze out
as asymptotic states.

Here, we perform a dynamical computation of hadron production from the decay of
a deconfined state with a spontaneously broken global $Z(3)$ symmetry without
assuming the formation of an equilibrated hadron gas {\em a priori}.
For simplicity, we restrict ourselves to the pseudoscalar
meson octet, i.e.\ the pions, the kaons, and the eta. At the end of the
paper, we also
speculate about possible differences in the ``statistical hadronization'' in
small versus large systems.

We begin by briefly reviewing the model for the QGP suggested recently by
Pisarski~\cite{Pisarski:2000eq}, which we shall adopt here to compute
hadron production.
The basic postulate of the so-called ``Polyakov loop model'' is that the
free energy of the deconfined state of QCD
is dominated by a condensate of gauge invariant Polyakov loops 
\be
\ell = \frac{1}{3} \; {\rm tr}\; {\cal P} 
\exp \left( i g \int^{1/T}_0 A_0(\vec{x},\tau) \, 
d\tau \right)~.
\ee
It is based on the observation from $SU(3)$ lattice gauge theory
that the Polyakov loop becomes light near
$T_c$~\cite{pol_mass,Karsch:2000ps}.
In other words, the effective potential for $\ell$ must be rather flat
for $T=T_c$. On the lattice, the mass of $\ell$ can be measured from the
two-point function with space-like separation. These two point functions in
the Polyakov loop model are very different from
those of ordinary perturbation theory~\cite{Dumitru:2001xa}.
The condensate is then characterized by a length scale $\xi=1/m_\ell(T)$.

In order to study hadronization, we couple the Polyakov loop to the
$SU(3)\times SU(3)$ linear sigma model and make qualitative
predictions for the resulting relative multiplicities of kaons and
etas to pions.  The results of this study have implications for
understanding the flavor composition in the final state of high-energy
inelastic reactions.  For simplicity, we neglect the effects of
nonzero baryon charge, isospin charge, electric charge, etc.\ in the
present analysis. This should be a reasonable first approximation for
particles produced in the central region of very high energy (and high
multiplicity) inelastic processes. In other cases, one may extrapolate
the measured hadronic multiplicities to vanishing
charges~\cite{kabana}, for example, so as to allow for a better
comparison of $e^+e^-$, $pp$, and nucleus-nucleus collisions.

Our model
has static, thermodynamic properties which are consistent with lattice
data for the pressure, the energy density, and the two-point function (i.e.\
the mass) of the Polyakov loop.
With the simplest possible $Z(3)$ symmetric potential described below,
one can not only fit the above observables but actually relate
their behavior, for example the pressure above
$T_c$ to the electric screening mass~\cite{Dumitru:2001xa}.
One also obtains testable predictions for the ratio of the masses
of the real and imaginary parts of the Polyakov loop~\cite{Dumitru:2001xa}.

\begin{figure}[htp]
\centerline{\hbox{\epsfig{figure=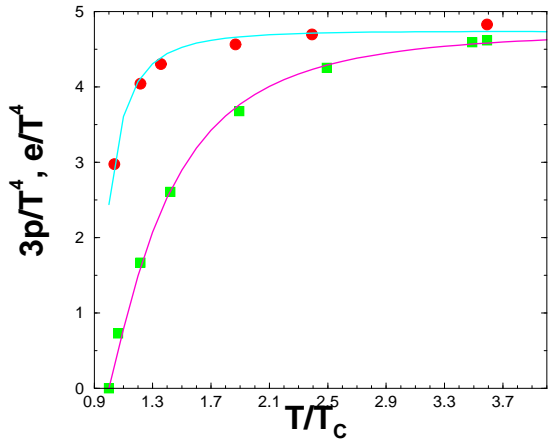,height=6cm}}}
\caption{The energy density and the pressure for SU(3) pure-gauge theory,
and our fit (see text).}  \label{fig:pot1}
\end{figure}
For three colors, we take the potential for $\ell$
to be
\begin{equation} \label{ellpot}
{{{\cal V}(\ell)=\left ( -\frac{b_2}{2}
|\ell|^2-\frac{b_3}{6}(\ell^3+(\ell^*)^3)+\frac{1}{4}
(|\ell|^2)^2\right )b_4T^4}},
\end{equation}
which is invariant under global $Z(3)$ transformations
\cite{Pisarski:2000eq,Dumitru:2001in}.
The form of ${\cal V}(\ell)$
is dictated by symmetry principles~\cite{Pisarski:2000eq,svetit}.  The center 
of $SU(3)$, $Z(3)$, is broken at high $T$ where 
the global minimum of ${\cal V}(\ell)$,
$\ell_0(T)\to 1$, and restored at low $T$ where $\ell_0(T)\to 0$. 

We assume that quarks are not important for understanding the form of
the effective potential for QCD.  This is motivated by the results of
lattice simulations for which $P/P_{\rm ideal}$ as a function of
$T/T_c$ is remarkably insensitive to the number of flavors, $N_f= 0$,
$2$, $3$~\cite{Karsch:2000ps}.  Thus, we neglect terms linear in
$\ell$, which exist in QCD with quarks.  Such terms would change the
weakly first-order transition (for $N_f=0$) which is predicted by 
eq.\ (\ref{ellpot}) into a crossover.
Nevertheless, the fact that the free energy is small just below $T_c$ means
that their numerical effects must be small.  We emphasize that the
results to be described below are {\em not\/} driven by the order of
the phase transition and would not change in any significant way by
the inclusion of such small linear terms.  All that matters is that
the $\ell$ field becomes rather light at $T_c$, i.e.\ that the
transition is {\em nearly} second-order.

The coefficients $b_2(T)$, $b_3$, and $b_4$ can be chosen to reproduce
lattice data for the pressure and energy density of the pure glue
theory for $T \ge T_c$~\cite{su3p}.  For example, one may chose $b_4\approx
0.6061$, $b_3=2.0$, and
$b_2(T)=(1-1.11/x)(1+0.265/x)^2(1+0.300/x)^3-0.487$
\cite{Dumitru:2001in}, where $x\equiv T/T_c$.  For this
parameterization, one obtains the pressure and energy density
$e=Tdp/dT-p$ as shown in Fig.~\ref{fig:pot1}.
The above parameterization does not yet incorporate the normalization
condition that $\ell_0(T)\to1$ for $T\to\infty$. Rather, the
expectation value for $\ell$ at high temperature approaches $r\equiv
b_3/2+\sqrt{b_3^2+4b_2(\infty)}/2$. We thus rescale the
field and the coefficients as $\ell\to\ell/r$, $b_2(T)\to b_2(T)/r^2$,
$b_3\to b_3/r$, and $b_4\to b_4 r^4$. This ensures the proper
normalization for the Polyakov loop. The rescaling does not, of course,
change the pressure and the energy density.

\begin{figure}[htp]
\centerline{\hbox{\epsfig{figure=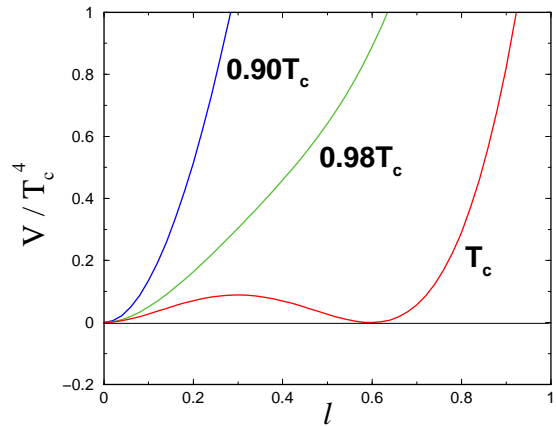,height=6cm}}}
\caption{The potential for the real part of $\ell$ at
three different temperatures. }
\label{fig:pot2}
\end{figure}
For $N_f=3$, we rescale $b_4$ by a factor of $(47+\frac{1}{2})/16$,
corresponding to the increase in the number of degrees of freedom. In
the spirit of mean field theory, the same constant values of $b_3$ and
$b_4$ are employed for $T < T_c$.  In this domain,
$b_2(T)$ can be set using the
string tension $\sigma(T)$ from the lattice~\cite{Karsch:2000ps}:
$g^2/2N_c\,\,\partial^2 {\cal V}/\partial\ell^2 = - g^2 b_2(T) b_4 T^4/2N_c=
\sigma^2(T)$. The factor in front of ${\cal V}''$ arises from the normalization
of the kinetic term for $\ell$, see below. Taking $N_c/g^2=1$, this leads to
$b_2(T)\approx -0.66 \, \sigma^2(T)/T^4$. 

The potential in eq.~(\ref{ellpot}), which follows from the above-mentioned
fit to the lattice data, changes extremely rapidly below
$T_c$~\cite{Dumitru:2001in}, see Fig.~\ref{fig:pot2}.
This rapid change in the effective potential with temperature
has important implications for the dynamical evolution of the confinement
transition~\cite{Scavenius:2001pa}, and for the relative hadron multiplicities
from the decay of the condensate for $\ell$. The hadron production
temperature $T_h$
can not vary much between events because it is impossible for the
expectation value for $\ell$ to stay large below $T_c$; rather, the field is
in an unstable configuration and must roll down towards the minimum of the
potential. (In turn, if the effective potential was slowly varying about $T_c$,
one would expect a broad range of hadronization temperatures.)
Also, $T_h$ must be very close to $T_c$, which in turn implies that
the $K/\pi$ ratio is well below unity. As we shall see below, if we 
chose $T_h$ by hand to be some $10\%$ or more below $T_c$
(disregarding the fact that from Fig.~\ref{fig:pot2} this scenario
appears unphysical), then phase-space for kaon production opens up and leads
to essentially equal number of kaons and pions in the final state, contrary
to experimental results. Thus,
the question as to whether it is some feature of QCD that the hadron production
temperature is so close to the confinement temperature $T_c$~\cite{specht}
is answered by the behavior of the potential for the Polyakov loop which we
deduce from the lattice data.

To complete the effective theory, we add a kinetic term for $\ell$ and 
couple it to the octet of pseudo-Goldstone excitations.
The coefficient of the kinetic term for $\ell$ is set by the
fluctuations of the $SU(3)$ Wilson lines in space. Perturbative corrections
to that coefficient are small~\cite{Wirstam:2001ka}.  We assume here a
Lorentz invariant form~\cite{Dumitru:2001in,Scavenius:2001pa}.
The coupling of $\ell$
to the $SU_R(3)\times SU_L(3)$ linear sigma model is given by 
\begin{equation}
{{{\cal L} \; = \; {\cal L}_\Phi + 
\frac{N_c}{g^2}
|\partial_\mu \ell|^2 T^2 - {\cal V}(\ell) - h^2 |\ell|^2 T^2
\, \mbox{Tr}\, \left(\Phi \Phi^\dagger\right)}}
\label{ec}
\end{equation}
where ${\cal L}_\Phi$ is the Lagrangian for the linear sigma model 
given below and $\Phi$ is the chiral field.
The coupling $h^2 \simeq 20$ between
the chiral field and $\ell$ is chosen to reproduce 
$m_\pi(T)$~\cite{Gavai:2000mx}.  
Through this interaction term, the Polyakov loop condensate drives the chiral
phase transition.

The Lagrangian of the $SU(3)_R \times SU(3)_L$ 
linear sigma model is given by \cite{SU3}
\bea \label{Lphi}
{\cal L}_\Phi &=& 
{\rm Tr}  \left( \partial_{\mu} \Phi^{\dagger} 
\partial^{\mu} \Phi 
-  m^2 \, \Phi^{\dagger} 
\Phi \right) - 
\lambda_{1} \left[ {\rm Tr}  \left( \Phi^{\dagger}
 \Phi  \right) \right]^{2} \\ \nonumber 
&-& \lambda_{2} {\rm Tr}  \left( \Phi^{\dagger} 
 \Phi  \right)^{2} 
+ c \left[ {\rm Det} \left( \Phi \right) + 
{\rm Det}  \left( \Phi^{\dagger} \right) \right]  \\ \nonumber
&+& {\rm Tr} \left[H  (\Phi + \Phi^{\dagger})\right] \,\, . 
\eea
$\Phi$ is a complex $3 \times 3$ matrix parameterizing the
scalar and pseudoscalar meson nonets,
\be
\Phi =\frac{\hat{\lambda}_{a}}{2} \, \phi_{a} =   
 \frac{\hat{\lambda}_{a}}{2} \, (\sigma_{a} + 
        i \pi_{a})\,\, , \label{defphi} 
\ee
where $\hat{\lambda}_{a}$ are the Gell-Mann matrices with 
$\hat{\lambda}_{0} = \sqrt{\frac{2}{3}} \, {\bf 1}$. 
The $\sigma_{a}$ fields are members of the 
scalar nonet and the $\pi_{a}$ 
fields are members of the pseudoscalar nonet.
Here, $\pi^{\pm} \equiv (\pi_{1} \pm i \, \pi_{2})/\sqrt{2}$ 
and $\pi^{0} \equiv \pi_{3}$ are the charged and neutral 
pions, respectively.  $K^{\pm} \equiv (\pi_{4} \pm i \, 
\pi_{5})/\sqrt{2}$, $K^{0} \equiv (\pi_{6} + i \, \pi_{7})/\sqrt{2}$,
and $\bar{K}^{0} \equiv (\pi_{6} - i \, \pi_{7})/\sqrt{2}$ are 
the kaons.  The $\eta$ and the $\eta'$ mesons are admixtures 
of $\pi_0$ and $\pi_8$. The classification of the scalar 
nonet is not important for our purposes as discussed below. 

The $3 \times 3$ matrix $H$ is chosen to 
reproduce the mass splitting of the strange and 
nonstrange quarks with $H = H_{a}{\hat{\lambda}_{a}}/{2}$ and 
where $H_{a}$ are nine external fields.  The determinant terms
reproduce the effects of the $U(1)_A$ anomaly in the QCD vacuum
\cite{tHooft} by preserving the $SU(3)_R \times SU(3)_L \cong
SU(3)_{V} \times SU(3)_{A}$ symmetry while explicitly breaking
the $U(1)_{A}$ symmetry.
Symmetry breaking gives the $\Phi$ field a vacuum 
expectation value, $\langle \Phi \rangle \equiv 
\bar{\sigma}_{a} {\hat{\lambda}_{a}}/{2}$.
The $\bar{\sigma}_0$
and $\bar{\sigma}_8$ are admixtures of the 
strange and nonstrange quark--antiquark condensates.
From the PCAC relations, they are given by 
$\bar{\sigma}_0=\left(f_\pi+2f_K\right)/\sqrt{6}=130$~MeV and 
$\bar{\sigma}_8=2\left(f_\pi-f_K\right)/\sqrt{3}=-23.8$~MeV.
Here, for simplicity
we freeze $\bar{\sigma}_0$, $\bar{\sigma}_8$ at their vacuum values,
i.e.\ we assume that dynamical relaxation of the classical condensates does
not contribute much to particle production or to the tadpole self-energies
of the $\pi$, $K$, $\eta$ quantum fields. From the numerical values of
$\bar{\sigma}_0$, $\bar{\sigma}_8$ it is clear that the energy density of the
chiral condensate is too low to contribute much to the total hadron
multiplicities. The free energy of the deconfined state is largely due to
the potential~(\ref{ellpot}) for the Polyakov loop $\ell$ which dominates
hadron production at $T_c$~\cite{Dumitru:2001in}.

The parameterization of the coupling constants is given in detail in
Ref.\ \cite{SU3} and is chosen to reproduce the masses and mixing
angles of the mesons at zero temperature.  Here, we take $m^{2} =
(342\, {\rm MeV})^2$, $\lambda_{1} = 1.4$ and $\lambda_{2} =
46$.  The explicit symmetry breaking terms are given by $H_0 =
(286\, {\rm MeV})^3$, $H_8 = -(311 \, {\rm MeV})^3 $ and $c =
4808\, {\rm MeV}$.  

To simplify the solution of the equations of motion for $\ell$ and the 
mesonic excitations, we do not consider the excitations of the 
scalar nonet or the $\eta'$, all of which
have masses either near or above 1~GeV. 
We perform the standard decomposition
of the meson quantum fields in terms of creation and annihilation operators
times adiabatic mode functions ${\pi}^k_a(t)$. 
Here, $k$ is the mode number. 
To make the interaction quadratic in the meson fields, we employ a 
mean field factorization which respects the flavor-$SU(3)$
symmetry when fluctuations are large,
\be
\pi_a^2(t,\vec{x}) \rightarrow \langle \pi_a^2(t)\rangle~.
\ee
With these approximations, the interaction part becomes
\bea
-{\cal L}_{\rm int} &=& \frac{1}{2}
   \left(\lambda_1+\frac{1}{2}\lambda_2\right) \biggl(
  \pi^2 + K^2 + \eta^2 \biggr)\langle\phi^2\rangle\nonumber\\
 &+& \frac{1}{2}h^2 T^2 |\ell|^2 \biggl(
   \pi^2+K^2+\eta^2\biggr)~,
\label{L_int}
\eea
where we introduced the short-hand notation
\be
\langle\phi^2\rangle\equiv
3 \langle\pi^2\rangle + 4 \langle K^2\rangle
   + \langle\eta^2\rangle
\ee
for the sum of the fluctuations. The prefactor of each term reflects the
internal isospin degrees of freedom. 
Assuming that the up and down quarks are mass-degenerate, 
the equations of motion for the mesonic
mode functions in Minkowski space read
\bea
 (\partial^2/\partial t^2 &+& k^2)\pi^k\nonumber\\
&=&\bigg[
   -m^2 - \lambda_1 \left( \langle\phi^2\rangle
           + \bar{\sigma}_0^2 + \bar{\sigma}_8^2\right)
  \nonumber \\
  &-& \lambda_2 \left( \frac{\langle\phi^2\rangle}{2}
        + \frac{\bar{\sigma}_0^2}{3}
        + \frac{\sqrt{2} \bar{\sigma}_0 \bar{\sigma}_8}{3} 
        + \frac{\bar{\sigma}_8^2}{6}\right)
        \nonumber \\ 
  &+&c \left( \frac{\bar{\sigma}_0}{\sqrt{6}} 
        - \frac{\bar{\sigma}_8}{\sqrt{3}}\right)
        -h^2 |\ell|^2 T^2 \bigg] \pi^k~,                      \label{pion}\\
 (\partial^2/\partial t^2 &+& k^2) K^k\nonumber\\
&=&\bigg[
  -m^2 - \lambda_1 \left( \langle\phi^2\rangle
        + \bar{\sigma}_0^2 + \bar{\sigma}_8^2\right)
  \nonumber\\
  &-& \lambda_2 \left( \frac{\langle\phi^2\rangle}{2}
  + \frac{\bar{\sigma}_0^2}{3}
    -\frac{\sqrt{2} \bar{\sigma}_0 \bar{\sigma}_8}{6} + 
    \frac{7 \bar{\sigma}_8^2}{6}\right)\nonumber\\
  &+&c \left( \frac{\sqrt{2} \bar{\sigma}_0 + \bar{\sigma}_8}{2\sqrt{3}}
      \right) - h^2 |\ell|^2 T^2 \bigg] K^k~,                 \label{kaon}\\ 
 (\partial^2/\partial t^2 &+& k^2) \eta^k\nonumber\\
&=&\bigg[
  -m^2 - \lambda_1 \left( \langle\phi^2\rangle
        + \bar{\sigma}_0^2 + \bar{\sigma}_8^2\right) 
    \nonumber\\
  &-&\lambda_2 \left(\frac{\langle\phi^2\rangle}{2}
     + \frac{\bar{\sigma}_0^2}{3}  
        - \frac{\sqrt{2} \bar{\sigma}_0 \bar{\sigma}_8}{3} + 
        \frac{\bar{\sigma}_8^2}{2} \right) \nonumber\\
  &+& c \left( \frac{\bar{\sigma}_0}{\sqrt{6}}+ 
      \frac{\bar{\sigma}_8}{\sqrt{3}}\right) - h^2 |\ell|^2 T^2
                                                    \bigg]\eta^k \label{p8}~.
\eea
The equations of motion are now $N_k$ (coupled) equations for the mode
functions, where $N_k$ is the number of modes. In the actual
computations we propagated $N_k=1000$ modes in parallel, taking a
mode-spacing of $\Delta_k=f_\pi/100$.  We checked that the
observables, like the total number of produced pions and kaons, were
stable under reasonable variations of $N_k$. When discussing relative
abundances of $\pi$, $K$, $\eta$ below, we always refer to the
sum of the occupation numbers of all field modes.
For the Polyakov loop, only the zero-mode (condensate) is considered
here. The equation of motion for $\ell$ is
\be
\frac{N_c}{g^2} T^2 \frac{\partial^2\ell}{\partial t^2} = -
\frac{\delta {\cal V}(\ell,\ell^*)}{\delta \ell^*} - \frac{h^2}{2}
T^2 \langle\phi^2\rangle \ell~.
\ee
This equation follows from~(\ref{L_int}) when fluctuations of the $\ell$ field
are disregarded, as in the present qualitative analysis. Such an approximation
is valid for short times, when the energy density is dominated by
the zero-mode of the Polyakov loop and the contributions from fluctuations
of $\ell$ and of the chiral fields (the produced mesons) are not large.

The fluctuations of the chiral fields at any given instant are computed
from their mode functions,
\bea
\langle\pi_a^2(t)\rangle &=&
\int \frac{d^3k}{(2\pi)^3} |\pi_a^k(t)|^2 \nonumber\\
 & & \hspace{-1.3cm} =
\int \frac{d^3k}{(2\pi)^3} 
\left\{ \frac{{\cal N}_k(t)}{\Omega_k(t)}
+\frac{1}{2\Omega_k(t)} - \frac{1}{2\Omega_k(t=0)} \right\}~.
\eea
The contribution from the vacuum fluctuations diverges.
The integral is regularized using the UV cutoff $\Lambda=N_k\Delta_k=10f_\pi
\simeq1$~GeV, which is a sensible value for our effective theory.
In practice, the dependence on this cutoff was found to be weak; 
if $\Lambda$ exceeds all other mass scales, the dependence on $\Lambda$
is only logarithmic~\cite{Dumitru:2001in,john_dirk}. 

In the above equation, ${\cal N}_k$ denotes the occupation number of 
a mode with given isospin and with momentum $k$,
subtracted for vacuum fluctuations,
\be
{\cal N}_k = \frac{\Omega_k}{2} \left(\frac{|\dot{{\pi}}^k_a|^2}{\Omega_k^2} 
+ |\pi^k_a|^2 \right) - \frac{1}{2} \; ,
\ee
and $\Omega_k=\sqrt{k^2+m^2_{\pi_a}+\Pi^2_{\pi_a}}$. $\Pi_{\pi_a}$
is the tadpole self-energy of meson field $\pi_a$. With the approximations
that lead to the interaction~(\ref{L_int}), it is obtained by taking two
functional derivatives of ${\cal L}_{\rm int}$ with respect to
$\pi_a$, which is trivial since ${\cal L}_{\rm int}$ is quadratic in the
dynamical fields. The expressions for $m^2_{\pi_a}+\Pi^2_{\pi_a}$ can simply
be read off from the r.h.s.\ of eqs.~(\ref{pion},\ref{kaon},\ref{p8}).

For the initial condition at $t=0$, we consider the case that only adiabatic
quantum fluctuations of the meson fields exist, but no on-shell particles,
i.e.\
\be
\Omega_k(0) \pi^k_a(0) = i \dot\pi^k_a(0)
= \sqrt{\Omega_k(0)/2}~.
\ee

To calculate the particle production dynamically, we make some
simplifications. The dynamical picture is that of a quench to a
temperature $T_h$, where the condensate for $\ell$ decays into
hadrons. ($T_h$ is not to be confused with the decoupling temperature of the
chiral field $\Phi$ from the $\ell$ field.)
The quench approximation seems reasonable as the effective
potential changes very rapidly around $T_c$. Dynamical simulations of
the long wavelength modes on a lattice show that the $\ell$-condensate
can be quenched to perhaps a few percent below
$T_c$~\cite{Scavenius:2001pa}.  Quenching to lower temperature is
hardly possible, since the $\ell$ field eventually
just ``rolls down'' towards the confined minimum.
This is also evident from the rapidly steepening potential in 
Fig.\ \ref{fig:pot2}.

\begin{figure}[htp]
\centerline{\hbox{\epsfig{figure=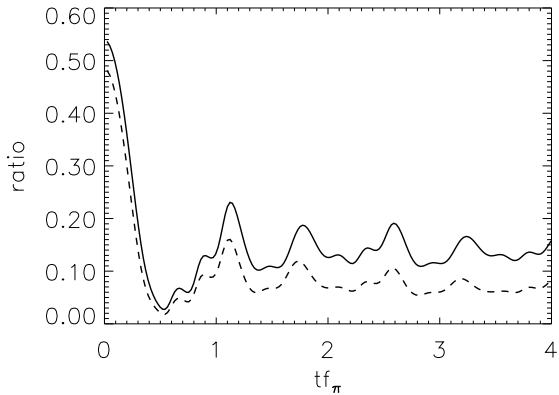,height=6cm}}}
\caption{The $K^{ch}/\pi^{ch}$ (solid line)
and the $\eta/\pi^0$ (dashed line) ratios as functions of time.}
\label{K2pi1}
\end{figure}

In Fig.\ \ref{K2pi1}, we show the result of such a computation, using
a hadron production temperature of $T_h=0.96 T_c$ and an initial
value for the expectation value of the Polyakov loop of
$\ell(t=0)=0.7$.  The $\ell$-field ``rolls down'' towards the minimum
at $\ell=0$, and thereby the energy of the deconfined state is
converted into physical hadronic states. Initially, as few hadrons
have been produced, the relative number of kaons and pions fluctuates.
However, it quickly settles to about
$K/\pi\approx 0.15$.  The relative suppression of kaon production is
due to its larger mass. At $T_h=0.96
T_c$, the mass for the real part of $\ell$ (defined as the curvature
of the potential about $\ell=0$), $m_\ell \approx 400$~MeV, is 
between the vacuum masses of the pion and the kaon, hence the phase space
for decay into kaons is smaller than for the decay into pions.

The nearly perfect agreement of our result on the $K/\pi$ ratio with
experimental data from the
CERN-SPS~\cite{KpiSPS} and BNL-RHIC~\cite{KpiRHIC} accelerators is certainly
a numerical coincidence. For example, variations of the coupling constants
and in particular of $T_h$ will certainly change the result somewhat. Since
this is not our major concern here, we refrain from a more detailed discussion.
Rather, the main point is that from the behavior of the effective Lagrangian
for the Polyakov loop depicted in Fig.~\ref{fig:pot2} one is led to expect
that the $\ell$ field decays into hadrons at $T_h$ very close to $T_c$.
At such temperature then,
$m_\pi < m_\ell(T_h) < m_K$ and so the available phase
space leads to much fewer kaons than pions.

\begin{figure}[htp]
\centerline{\hbox{\epsfig{figure=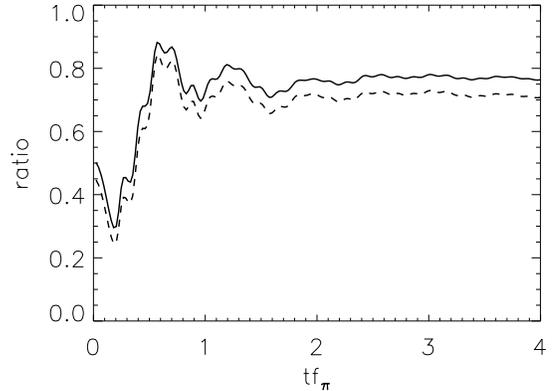,height=6cm}}}
\caption{The $K^{ch}/\pi^{ch}$ (solid line)
and the $\eta/\pi^0$ (dashed line) ratios as functions of time
when $T/T_c=0.87$.}
\label{K2pi2}
\end{figure}

To illustrate this fact, we repeat the computation, lowering $T_h$ by hand.
While this is unrealistic because it assumes that the condensate for
$\ell$ retains a large expectation value even below $T_c$, it shows
that as $T_h$ is lowered, the decay of
the condensate for $\ell$ leaves us with essentially equal numbers of
kaons and pions. This is due to the increase of $m_\ell$, which makes
phase-space restrictions less relevant.
Fig.~\ref{K2pi2} confirms this expectation. If the hadrons were
produced at $T=0.87T_c$, where $m_\ell\approx 740$~MeV
is above the kaon vacuum mass, we obtain $K/\pi\approx 0.8$.

The theoretical expectation that the decay of the condensate for $\ell$ into
hadrons occurs close to $T_c$, which
was based on the rapid variation of the effective potential for $\ell$ from
Fig.~\ref{fig:pot2}~\cite{Dumitru:2001in,Scavenius:2001pa}, is thus perfectly
consistent with the constraints from experimental data on the $K/\pi$ ratio
in high-energy heavy-ion scattering at the CERN-SPS~\cite{KpiSPS}
and BNL-RHIC~\cite{KpiRHIC}.
What would perhaps be even more interesting from the point of view of
learning about the presence of fundamental QCD scales in the hadronization
process is to look for situations where our calculation {\em breaks down}.
Fortunately, there indeed seems to be an experimentally accessible regime
where this can be tested, namely high-energy,
high-multiplicity $pp$ inelastic reactions. 

At given energy $\sqrt{s}/A$, and at central rapidity,
the $K/\pi$ ratio in heavy-ion collisions
approaches that from $pp$ (or $p\bar{p}$) reactions at
the same $\sqrt{s}$ as the energy increases~\cite{kabana,redlich}.
A natural question to ask is whether indeed they
asymptotically approach the same limit. In fact, one may even compare
central heavy-ion collisions at, say, BNL-RHIC energy to high-multiplicity
$pp$ (or $p\bar{p}$) at much higher energies, where the charged multiplicity
per participant and the energy density at central rapidity are comparable
to those from heavy-ion collisions,
and where one may very well produce a transient QGP state.
In the absence of a length scale $1/m_\ell$,
one would naively expect that eventually, at sufficiently high energy,
the $K/\pi$ ratio from high-multiplicity $pp$ approaches that from central
heavy-ion collisions at BNL-RHIC energy (also if high-multiplicity fluctuations
in $pp$ are due to jets).

On the other hand, if the free energy of QCD around $T_c$ is dominated
by a condensate for the Polyakov loop, hadron production is governed by a
scale $m_\ell$ as explained above. Thus, hadronization from the decay of the
condensate is associated with a length scale $1/m_\ell$. It appears quite
reasonable to assume that it can only dominate hadron
production in systems of size greater than 
$1/m_\ell$. (In small systems, on the other
hand, fluctuations dominate.) With $m_\ell$ somewhere between
$m_\pi$ and $m_K$, as given by the potential for
$\ell$ from Fig.~\ref{fig:pot2},
this condition may not hold for $pp$ reactions, {\em regardless of how
high an energy density is achieved}. Note that we are {\em not} referring here
to thermodynamical effects like for example canonical suppression
due to exact conservation of strangeness (which becomes less important for very
high energy and high-multiplicity $pp$, when there is more than one charged
kaon pair per unit of rapidity). We are referring to whether or not one can
probe, in principle, the deconfined phase of the non-abelian $SU(3)$
gauge theory in
a system of linear dimension $1/m_\ell$ (or less), at a temperature
$T\sim T_c$.

Consider pure gauge theory with three colors, for which good lattice data
exists. In $d$ Euclidean dimensions,
the two-point function for the real part of the Polyakov
loop behaves as
\bea
\langle \ell_r(r) \ell_r(0)\rangle - \langle \ell_r\rangle^2 &\sim &
\int d^d k \frac{e^{i k\cdot r}}{k^2+m_\ell^2} \nonumber\\
&\sim&
\left(\frac{m_\ell}{r}\right)^{(d/2-1)} K_{d/2-1} (m_\ell\, r)~.
\eea
The correlation function for the imaginary part of $\ell$ falls off more
rapidly, corresponding to a larger mass for $\ell_i$~\cite{Dumitru:2001xa}.
For $d=3$, $K_{1/2} (m_\ell\, r) \sim 1/\sqrt{m_\ell\, r}
\exp(m_\ell\, r)$, hence
\be
\langle \ell_r(r) \ell_r(0)\rangle - \langle \ell_r\rangle^2 \sim 
\frac{1}{r} \; e^{-m_\ell\, r}~.  
\label{2pointf}
\ee
When the size of the
system $r\lton 1/m_\ell$, the correlation function~(\ref{2pointf}) exhibits
a power-law behavior 
\be
\langle \ell_r(r) \ell_r(0)\rangle - \langle \ell_r\rangle^2 \sim 
\frac{1}{r}~,
\label{2pointf_small_r}
\ee
for all ``accessible'' distances. The $r-$dependence of the subtracted
correlation
function~(\ref{2pointf_small_r}) is in fact that for an {\em abelian} pure
gauge theory (photons), namely just the Coulomb potential. 
The fact that the Polyakov loop aquires a mass $m_\ell$, which is the feature
of the non-abelian gauge theory, can only be probed on distance scales
$r\gton 1/m_\ell$.
Our computation of hadron production applies to systems larger than
$1/m_\ell$, when the line-line correlation function falls off exponentially and
when the free energy is dominated by the condensate for $\ell$. Qualitatively,
hadron production about $T_c$ in systems of size $1/m_\ell$
might differ from our model.

Data on the $K/\pi$ ratio at central rapidity from $p\bar{p}$ reactions
at $\sqrt{s}=1800$~GeV at the Fermilab-Tevatron have been analysed as a
function of charged multiplicity~\cite{E735a,E735b}.
For the highest-multiplicity events,
corresponding to $dN_{ch}/d\eta\approx25$ the initial energy density at
proper time $\tau_0=1$~fm/c was estimated to be 6~GeV/fm$^3$, comparable to the
average value from central Au+Au collisions at BNL-RHIC energy $\sqrt{s}/A=
100-200$~GeV~\cite{KpiRHIC}. A first analysis~\cite{E735a} seemed to indicate
that $K/\pi$ increases as a function of event multiplicity, from about 10\%
to about 14\%, nearly the value from Au+Au at RHIC. However, error bars for
the most inelastic events were large, and a second analysis~\cite{E735b}
did not confirm the rising $K/\pi$ ratio but showed it to be completely 
flat as a function of $dN_{ch}/d\eta$. A new analysis of this and other
observables with the colliding proton beams at RHIC would be most interesting.

In summary, we have discussed dynamical hadron production at the confinement
transition, 
which is realized here as the restoration of the global $Z(3)$ symmetry
for the Polyakov loop $\ell$.
Specifically, the proposal to understand QCD near
$T_c$ in terms of a condensate for 
$\ell$~\cite{Pisarski:2000eq,Dumitru:2001xa,Dumitru:2001in} predicts
that hadron production occurs very near $T_c$. This
follows from the rapid change of the effective potential for the Polyakov loop
about $T_c$, which we deduce from lattice
data on the finite-temperature string tension, pressure, and energy density.
From our effective Lagrangian, the mass of $\ell$ just below $T_c$ is
greater than the vacuum mass of the pion but less than that of the kaon.
Therefore, qualitatively one expects about an order of magnitude more
pions than kaons to be produced, as seen in experimental data from
CERN-SPS and BNL-RHIC.

The hadronization process is characterized by a length scale $1/m_\ell$, which
is on the order of $1/2-1$~fm, just below $T_c$. Thus, in small systems
hadronization may not occur from the decay of a condensate for
$\ell$, regardless of the initial energy density.
Differences in the $K/\pi$ ratio (and in other relative hadron multiplicities)
from high-multiplicity $pp$ to heavy-ion collisions at
the SPS or RHIC~\cite{becattini,vance}
could perhaps be due to a condensate scale $1/m_\ell$.
If such a point of view can be established experimentally,
it could provide further support for the formation of a deconfined phase
in collisions of heavy ions at high energies.\\[.5cm]
\noindent
{\bf Acknowledgements:} We thank Rob Pisarski for many helpful
discussions on the Polyakov loop model which eventually led us to this work.
We would also like to thank Nu Xu for his interest in $K/\pi$ in
high-multiplicity $pp$ reactions; Francesco Becattini, Dirk Rischke and Nu Xu
for comments related to this paper,
and Mark Gorenstein for discussions on statistical hadronization.
A.D.\ acknowledges support from DOE Grant DE-AC02-98CH10886.


\end{document}